\documentclass[%
 reprint,
 amsmath,amssymb,
 aps,
prb,
]{revtex4-2}

\usepackage{graphicx}
\usepackage{dcolumn}
\usepackage{bm}
\usepackage{empheq}
\usepackage{siunitx}
\usepackage{xcolor}

\usepackage{fixltx2e}

\begin{document}

\preprint{APS/123-QED}

\title{Piezoelectric microresonators for sensitive spin detection}

\author{Cecile Skoryna Kline}
\address{Quantum Engineering Technology Labs and Department of Electrical and Electronic Engineering, University of Bristol,
Woodland Road, Bristol BS8 1UB, United Kingdom}

\author{Jorge Monroy-Ruz}
\address{Quantum Engineering Technology Labs and Department of Electrical and Electronic Engineering, University of Bristol,
Woodland Road, Bristol BS8 1UB, United Kingdom}

\author{Krishna C. Balram}
\email{krishna.coimbatorebalram@bristol.ac.uk}
\address{Quantum Engineering Technology Labs and Department of Electrical and Electronic Engineering, University of Bristol,
Woodland Road, Bristol BS8 1UB, United Kingdom}

\date{\today}

\begin{abstract}
Piezoelectric microresonators are indispensable in wireless communications, and underpin radio frequency filtering in mobile phones. These devices are usually analyzed in the quasi-(electro)static regime with the magnetic field effectively ignored. On the other hand, at GHz frequencies and especially in piezoelectric devices exploiting strong dimensional confinement of acoustic fields, the surface magnetic fields ($B_{1}$) can be significant. This $B_1$ field, which oscillates at \qty{}{GHz} frequencies, but is confined to \qty{}{\um}-scale wavelengths provides a natural route to efficiently interface with nanoscale spin systems. We show through scaling arguments that $B_1{\propto}f^2$ for tightly focused acoustic fields at a given operation frequency $f$. We demonstrate the existence of these surface magnetic fields in a proof-of-principle experiment by showing excess power absorption at the focus of a surface acoustic wave (SAW), when a polished Yttrium-Iron-Garnet (YIG) sphere is positioned in the evanescent field, and the magnon resonance is tuned across the SAW transmission. Finally, we outline the prospects for sensitive spin detection using small mode volume piezoelectric microresonators, including the feasibility of electrical detection of single spins at cryogenic temperatures.
\end{abstract}

\maketitle

\section{Introduction}

Piezoelectric microresonators \cite{bhugra2017piezoelectric, morgan2010surface} have revolutionized wireless communication by enabling small form-factor, high performance radio frequency (RF) filters that can be compactly packaged into mobile phones. In addition, these devices have had a broad impact on areas ranging from sensing \cite{ballantine1996acoustic} to quantum communication \cite{balram2022piezoelectric}. A piezoelectric material enables conversion of RF electromagnetic fields into acoustic fields, which have wavelengths $\approx$ \qty{}{\um} at \qty{}{GHz} frequencies, $10^5$ smaller than the \qty{}{cm}-scale wavelengths of the RF fields. This deeply-subwavelength confinement is the key driver for the majority of applications involving piezoelectric devices.

The constitutive relations for a piezoelectric device \cite{IEEE} relate the stress ($\vec{T}$) induced by an applied electric field ($\vec{E}$). While this is strictly true for applied DC fields, the equations are extended to RF fields under the quasi-(electro)static approximation \cite{haus1989electromagnetic}, wherein the Poisson equation for electrostatics is substituted for Maxwell's equations for the electromagnetic field, and is solved along with the elastic wave equation to propagate acoustic fields in piezoelectric devices. The quasistatic approximation is usually justified because the deeply sub-wavelength confinement provided by the acoustic field ensures that the far-field electromagnetic radiation component is minimal. By definition, using this approximation forces the magnetic field to be strictly zero.

On the other hand, it is known that piezoelectric devices radiate electromagnetically \cite{mindlin1973electromagnetic} and that this radiation presents a limit on the achievable mechanical quality factor ($Q_m$) in piezoelectric resonators \cite{yong2009effects}. While the far-field radiation efficiency is low for the reasons outlined above, it can be significantly enhanced in a resonant geometry and provides significant size advantages in the design of very low frequency antennas \cite{kemp2019high}. In this work, we focus on the near-field (surface) component of this oscillating magnetic field ($B_1$), and ask if these $B_1$ fields can be exploited for improving the spin detection sensitivity of nanoscale electron spin resonance (ESR) experiments \cite{blank2017recent, morton2018storing}. Our aim is to apply ideas from cavity quantum electrodynamics (cQED) \cite{haroche2006exploring} to nanoscale spin systems \cite{soykal2010strong, schuetz2015universal}, with the key sensitivity enhancement being provided by the vastly reduced mode volume ($V_m$) of piezoelectric microresonators in comparison to their electromagnetic counterparts.

\section{Surface current density scaling in piezoelectric devices}

\begin{figure}
    \centering
    \includegraphics[width = 1.0 \columnwidth]{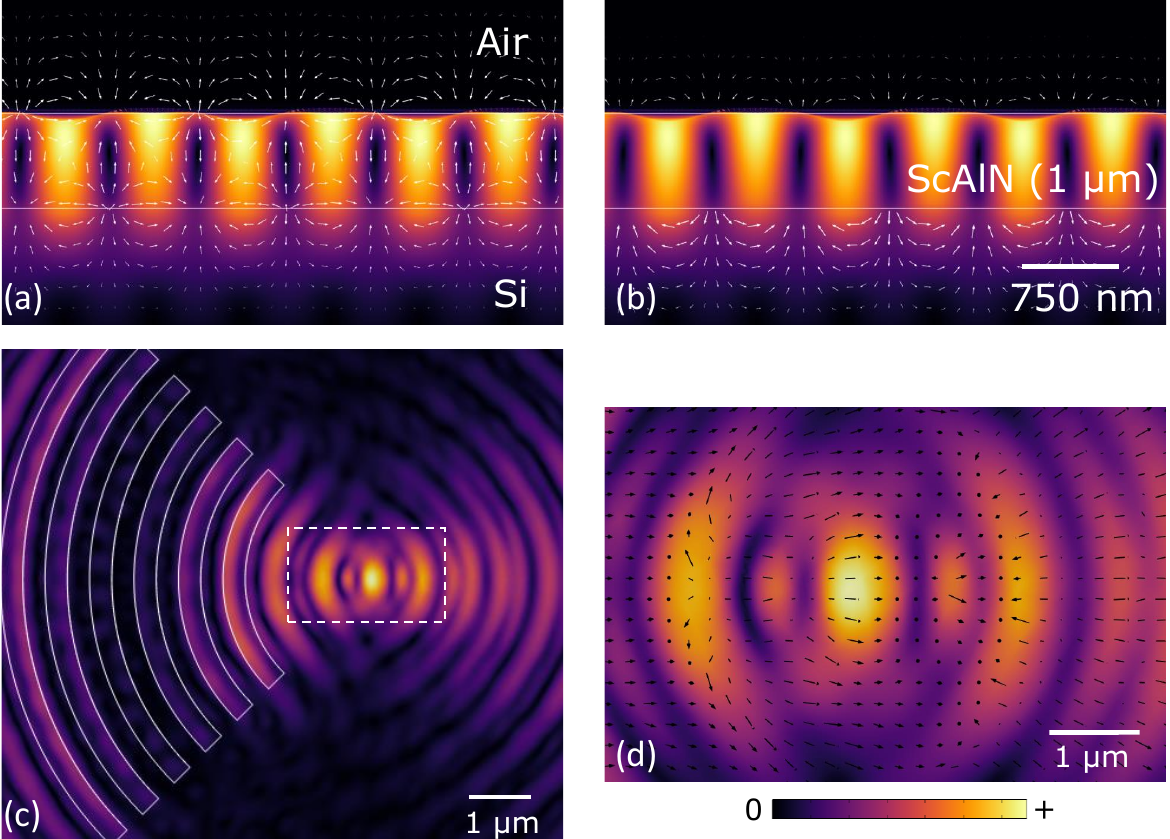}
    \caption{(a) FEM simulation showing the surface displacement of a surface acoustic wave propagating on a ScAlN on Si substrate. The accompanying $\approx$ \qty{2.9}{GHz} electric field is shown using an overlaid arrow plot. As can be seen there exists an evanescent field in the air whose orientation (helicity) is determined by spin momentum locking \cite{van2016universal}, (b) FEM simulation, same as (a) but showing the overlaid oscillating polarization current density using an arrow plot. This oscillating current density which exists at both the interfaces (air-ScAlN and ScAlN-Si) is responsible for the surface $\vec{B_1}$ fields we investigate in this work. (c) Focusing the SAW using curved electrodes in AlN (d) At the focus, the displacement, and corresponding evanescent field and current densities, are enhanced by the focusing ratio.}
    \label{SAW_surface_fields}
\end{figure}

In a piezoelectric material, the propagating acoustic displacement field is accompanied by a surface polarization ($\rho$). The evanescent electric fields ($\vec{E}$) generated at the material-air boundary curl as depicted in Fig.\ref{SAW_surface_fields}(a). The plot shows a finite element method (FEM) simulation of a surface acoustic wave (SAW) mode propagating on a Scandium Aluminum Nitride (ScAlN) on Si substrate. The accompanying electric fields are shown by an arrow plot. This curling of surface fields is one instance of the universal phenomenon of spin-momentum locking \cite{van2016universal} that applies to all evanescent fields (cf. Appendix \ref{App:SAW_SPP} for a discussion of the analogy between the surface fields in surface plasmons and surface acoustic waves). The accompanying surface magnetic field ($B_1$) orientation can be directly obtained from the ($\vec{E}$) fields, and in this case would form loops that come periodically into and out of the plane (shown schematically in Appendix \ref{App:SAW_SPP}). One can verify the $\vec{B_1}$ orientation by noting the direction of the surface polarization currents ($J = \frac{\partial\rho}{{\partial}t}$), plotted in Fig.\ref{SAW_surface_fields}(b) and invoking Ampere's law. For focusing acoustic field geometries as would be necessary for increasing the local field strength, the field orientation is shown in Fig.1(c), with a zoomed-in plot of the focus region shown in Fig.\ref{SAW_surface_fields}(d). We would like to note here that while the $\vec{E}$ field lines can be computed using an FEM solver like COMSOL, due to the quasistatic approximation being imposed on the solver, the magnetic field is strictly zero. Therefore, current FEM solvers cannot be used to visualize this ($B_1$) field directly from taking the curl of the $\vec{E}$ field. The results presented in Figs.\ref{SAW_surface_fields}(a-d) make certain simplifications. We model (Sc)AlN films using aluminum nitride's material parameters, and in Fig.\ref{SAW_surface_fields}(c,d), we simulate focusing on a thin AlN film to reduce the simulation's memory constraints.  

To estimate the scaling of this surface $B_1$ field, we therefore start with the oscillating surface current density ($J$, [\qty{}{\ampere\per\square\meter}]) instead \cite{ballantine1996acoustic} following prior work on SAW based sensors \cite{niemczyk1988acoustoelectric}:

\begin{equation}
    J^2 = 2K^2{\omega}k^2(\epsilon_c+\epsilon_s)P
    \label{eq:J_f}
    \end{equation}
    
where $K^2$ is the material's piezoelectric coefficient and represents the fraction of the acoustic wave energy that is stored in the electric field. $\omega$ is the wave frequency, $k = 2\pi/\lambda$ the wave vector, $\epsilon_{c,s}$ [\qty{}{\farad\per\meter}] represent the dielectric constants of the cladding (air) and substrate respectively, and $P$ is the power density of the acoustic wave expressed in power per unit beam width [\qty{}{\watt\per\meter}]. We can see that for a tightly focused acoustic beam with focus width $\approx\lambda$, $P{\propto}1/{\lambda}$, and hence $B_1^2{\sim}J^2{\propto}f^4$, or the surface field $B_1{\propto}f^2$. With $K^2$ = 0.05, $\lambda_a$= \qty{1}{\um}, $f$ = \qty{3}{GHz}, $\epsilon_c+\epsilon_s\approx$ 10, beam width at focus $\approx\lambda_a$ and $P_a$=\qty{1}{mW}, we expect a surface current density at the focus of $\approx$ \qty{81.15}{\mega\ampere\per\square\meter}. The scaling of $J$ with $\epsilon$ can be visualized directly in Fig.\ref{SAW_surface_fields}(b), where the current density at the (Sc)AlN-Si interface is stronger than at the (Sc)AlN-air interface. 

Assuming the current flows uniformly in a (semi-circular) loop of size $\approx\lambda_a$, one can roughly estimate the $B_1$ field as $B_1\approx{\mu_0}J\lambda_a/4$ = \qty{25.5}{\micro\tesla}, with $\mu_0$ the vacuum permeability. By trapping the acoustic field in wavelength scale microcavities \cite{msall2020focusing}, the surface current density and therefore the accompanying $B_1$ field can be enhanced by the cavity mechanical quality factor $Q_m$, with $Q_m\approx10^4$ feasible in crystalline media at ambient conditions \cite{bicer2023low}. Alternately, one can also estimate the spin-resonator single photon coupling strength $g$ in a cavity-QED framework \cite{haroche2006exploring, blank2017recent} with:
\begin{equation}
    g = \frac{\mu_B}{\hbar}\sqrt{\frac{2\mu_0\hbar\omega_0}{V_m}}
    \label{eqn:g}
    \end{equation}

where $\mu_B$ is the Bohr magneton, $\hbar$ the reduced Planck constant, $\mu_0$ vacuum permeability and $V_m$ is the cavity mode volume. Assuming an acoustic cavity mode with dimensions of $5\lambda_a^3$ at \qty{3}{GHz} with $\lambda_a$=\qty{1}{\um}, this gives us a coupling strength $g\approx$ $2\pi*$\qty{14}{kHz}. We would like to emphasize that we are interested in the near-field $B_1$ here. Due to the alternating signs of the surface current density, the far-field radiation is relatively weak.

Such piezoelectric approaches to enhancing spin detection sensitivity are in many ways complementary to superconducting cavity approaches which have recently achieved single spin sensitivity \cite{morton2018storing, wang2023single} at \qty{}{\milli\kelvin} temperatures. Both approaches rely on a Purcell-like \cite{haroche2006exploring} enhancement of the spin detection sensitivity which scales $\propto\sqrt{Q_c/V_c}$, where $Q_c$ is the quality factor and $V_c$ is the mode volume of the cavity. The superconducting approaches support high $Q$ factors ($>10^5$) and small magnetic mode volume ($\approx 10^{-12}\lambda_{RF}^3$) \cite{probst2017inductive} by exploiting lumped element inductor-capacitor ($LC$) cavity designs, where the $B_1$ field is strongly localized in close proximity to the inductor. Piezoelectric resonators, on the other hand provide moderate $Q_m$ ($\approx10^4$), but vastly reduced cavity mode volumes ($\approx10^{-15}\lambda_{RF}^3$) with the additional advantage of enabling experiments in ambient conditions, which makes it feasible to use this technique with chemical and biological samples which might deteriorate appreciably in cryogenic environments.

We would like to distinguish our work from previous results that have studied the interaction between acoustic fields and spin systems (both nanomagnets and magnetic thin films) that primarily exploit the magnetostrictive effect \cite{weiler2011elastically, li2017spin,  hatanaka2023phononic} or the strain field to perturb spin systems \cite{chen2019engineering, whiteley2019spin, maity2020coherent}. In this work, we instead focus on using piezoelectric devices for generating the oscillating magnetic field ($B_1$) directly and the methods developed here can be applied in principle to any spin resonance experiment. Given the field generated is on the surface and evanescent, physical contact between the sample and the spin system can be completely avoided, as discussed in the experiments below.

\section{Using spin resonance absorption to probe the evanescent $B_1$ field}

\begin{figure}
    \centering
    \includegraphics[width = 1.0 \columnwidth]{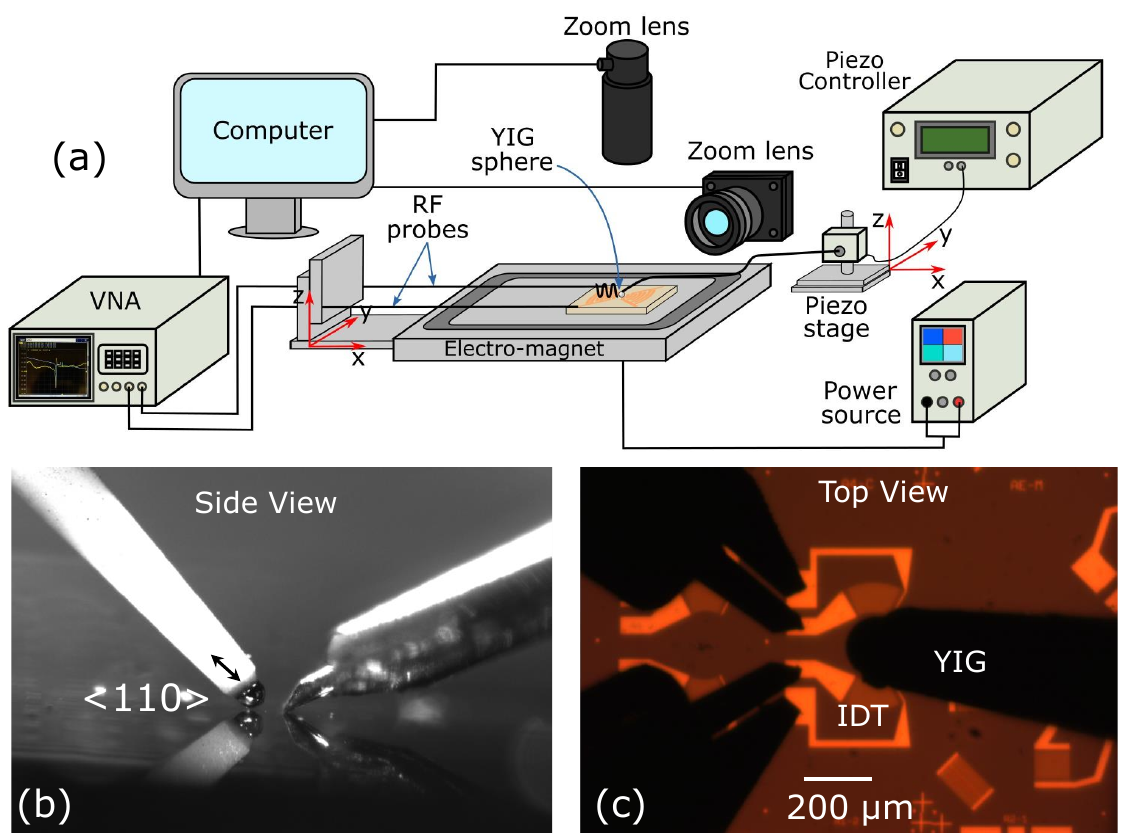}
    \caption{(a) Schematic of the experimental setup used to probe the $B_1$. Curved IDTs are used to launch and detect SAWs on a ScAlN on Si substrate. A polished YIG sphere is brought in proximity (non-contact) of the evanescent field at the focus of the SAW and the IDT transmission is monitored as the magnon mode is tuned across the SAW transmission resonance. The magnitude and phase of the transmission is monitored using a vector network analyzer and the magnon mode is tuned by a $B_z$ field applied using an electromagnet underneath the sample. (b) and (c) show images from the side and top view cameras of the experiment during operation.}
    \label{fig_setimg}
\end{figure} 

We perform spin resonance measurements by positioning a polished Yttrium-Iron-Garnet (YIG) sphere (diameter $\approx$ \qty{150}{\um}) in the evanescent field ($<\lambda_a\approx$ \qty{1}{\um}) of the SAW devices. YIG spheres support high-Q ($\approx 10^4$ at \qty{10}{GHz}) collective spin wave excitations (magnons), which have been used for a wide range of applications, ranging from tunable filters and low noise oscillators for wireless communication \cite{eliyahu2003tunable} to hybrid quantum transduction, wherein quantum states from superconducting qubits are mapped back and forth from the magnon modes \cite{tabuchi2015coherent}. By mounting the YIG sphere on a movable three-axis translation stage and positioning it in the evanescent field of the acoustic wave, we can spatially probe the surface $B_1$ field by monitoring the SAW transmission and looking for changes in the amplitude and phase response of the received SAW signal as the magnon mode frequency is tuned across the SAW resonance. 

The schematic of our experimental setup is illustrated in Fig.\ref{fig_setimg}(a). Fig.\ref{fig_setimg}(b,c) show the side and top view of the YIG sphere positioned at the centre of the SAW transmit-receive circuits as captured by the two zoom lenses indicated in Fig.\ref{fig_setimg}(a). SAWs are launched and detected using a vector network analyzer (VNA) by patterning a set of interdigitated transducers (IDT) \cite{morgan2010surface} on a piezoelectric (c-axis oriented) Sc\textsubscript{0.06}Al\textsubscript{0.94}N film deposited on a [100] oriented silicon substrate. We pattern IDTs with both straight fingers to launch quasi plane waves of sound, and curved fingers \cite{wu2005analysis} to focus the acoustic field down to a beam width of 1-5 $\lambda_a$. The curved IDT devices are designed as confocal transmit-receive pairs \cite{siddiqui2018lamb, msall2020focusing}. To achieve the highest spin detection sensitivity, focusing devices are necessary as they significantly enhance the local power density ($P$ in equation 1) by the focusing ratio ($\approx 20-50x$) and the strongest $B_1$ fields are therefore always generated at the focus. The size of the YIG sphere in our experiments presents two constraints. It requires us to separate the focusing IDTs by $\approx100-125\lambda_a$, which magnifies the effect of wave diffraction and the anisotropy of the underlying silicon substrate with the net result being a relatively low acoustic throughput.The second issue, which can be seen from Fig.\ref{fig_setimg}(c) is the difficulty of determining the precise acoustic focus location while aligning the YIG sphere for maximum SAW extinction.

\begin{figure*}
    \centering
    \includegraphics[width = 1.0 \textwidth]{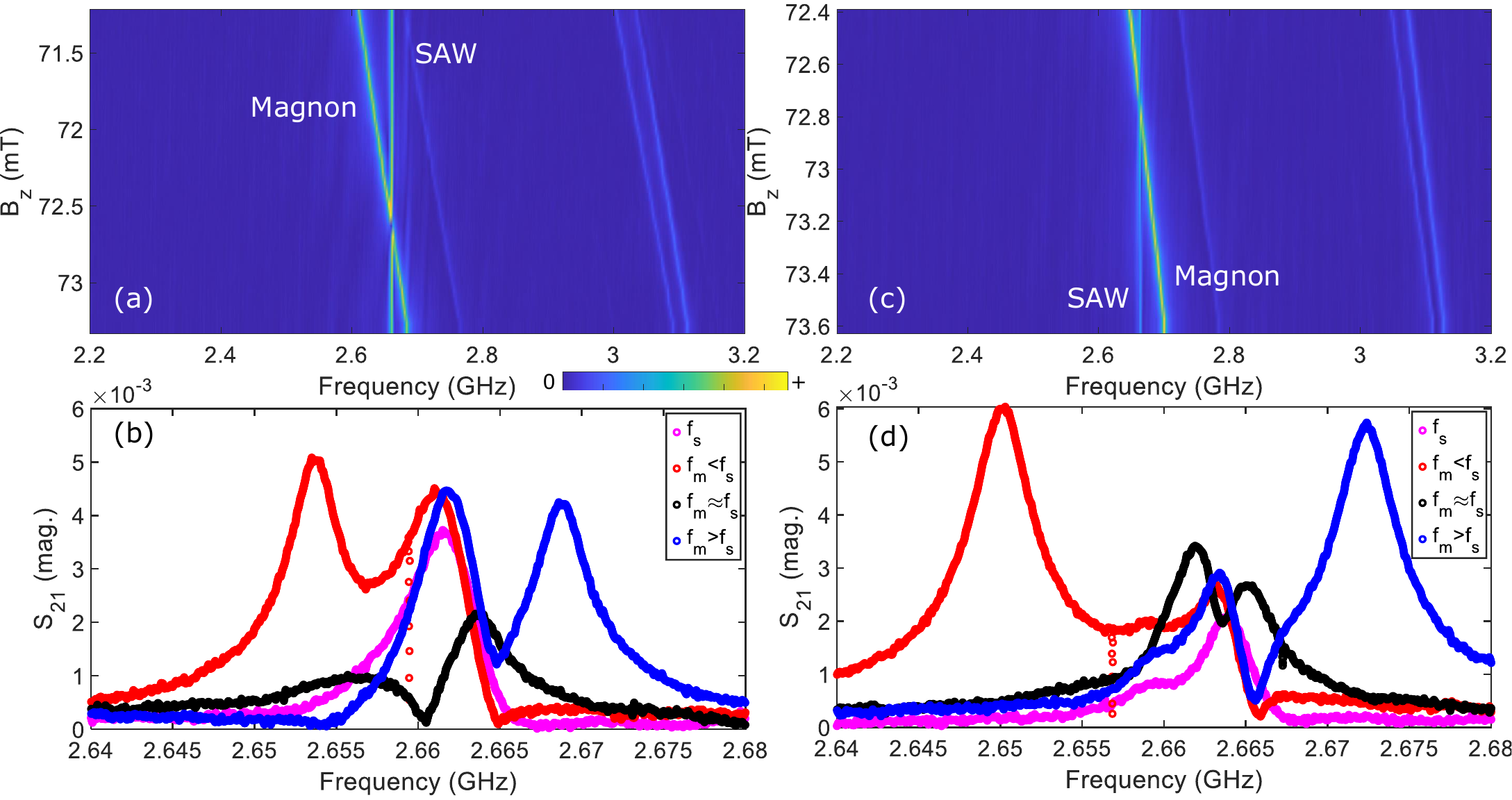}
    \caption{The SAW-magnon interaction is probed by monitoring the VNA transmission ($S_{21}$) as a function of $B_z$. The 2D colorplots in (a,c) plot $\lvert{S_{21}}\rvert$ as a function of frequency and $B_z$ for two different curved IDT devices with the same period, but different curvature. The SAW mode frequency is constant with $B_z$, whereas the magnon mode frequency shifts linearly with $B_z$. When the magnon mode frequency crosses the SAW transmission resonance, one expects an excess attenuation in the SAW transmission due to magnon mode excitation and dissipation. (b,d) show representative linecuts from (a,c) respectively for $f_m<f_s$ (red), $f_m{\approx}f_s$ (black), and $f_m>f_s$ (blue). The excess attenuation when $f_m{\approx}f_s$ is especially clear in (b), and the same trend can also be observed, albeit with lower magnitude in (d). The plotted data are time gated with a \qty{40}{ns}. Uncorrected datasets are available in Appendix \ref{App: Raw_data_sets}, and the bare IDT reflection and transmission spectra are plotted in Appendix \ref{App: IDT_spectra}.}
    \label{fig_data}
\end{figure*} 

The magnon frequency can be tuned by applying a static DC magnetic field ($B_z$) using an electromagnet, as shown in Fig.\ref{fig_setimg}(a). Due to the frequency of operation and the orientation (shown in Fig.\ref{fig_setimg}(b)), the YIG sphere is not saturated, and the magnon mode relaxes to lower frequencies with time. Therefore, there is a finite time offset between the setting of the voltage on the electromagnet and the data acquisition on the VNA, which would result in a lower effective $B_z$ due to relaxation. In principle, one can park the magnon mode on the higher frequency side of the SAW transmission and let the mode relax and down-shift in frequency across the SAW resonance, while recording VNA traces as a function of time. In practice, we found this method did not give us the resolution needed to capture the exact crossing between the magnon mode and the SAW. Therefore, we rely on incrementing the voltage in small steps and acquiring the data quickly to minimize the effects of mode relaxation.   

 As the $B_z$ field is tuned, ferromagnetic resonance (FMR) in the YIG sphere induces an excess absorption which shows up as a reduced transmission (attenuation) at the SAW  frequency. This excess absorption is maximized when the frequencies of the SAW and the magnon mode are identical. This is shown in Fig.\ref{fig_data}. Fig.\ref{fig_data}(a,c) correspond to datasets taken from two distinct curved IDT devices with the same period but different curvature. The colorbar on the 2D plot indicates the magnitude of the (normalized) transmitted signal ($S_{21}$) as a function of both $B_z$ and frequency.  As the plot shows, we observe a variety of magnon modes \cite{gloppe2019resonant} in our experiment, with varying coupling strengths and quality factors. Without mode imaging \cite{gloppe2019resonant}, it is hard to ascertain the mode symmetry from a pure microwave transmission experiment. Here, we focus on the mode that gives us the strongest signal and generically label it as magnon in Fig.\ref{fig_data}(a,c). As the 2D color plots show, the SAW frequency stays relatively independent of $B_z$, whereas the magnon mode linearly tunes to higher frequency with increasing $B_z$. When the magnon mode frequency ($f_m$) crosses the SAW mode ($f_s$), one can clearly see an excess absorption which is stronger in Fig.\ref{fig_data}(a) compared to the dataset in Fig.\ref{fig_data}(c). One can see this excess absorption effect more clearly by taking 1D cuts through the 2D data corresponding to the cases when $f_m<f_s$, $f_m{\approx}f_s$ and $f_m>f_s$. These 1D cuts are shown respectively in Fig.\ref{fig_data}(b) and (d), with the cases colored red, black and blue respectively. The background SAW transmission when the magnon mode is way off resonance ($f_m{\ll}f_s$, labelled $f_s$) is indicated in magenta. Especially, in Fig.3(b) the excess attenuation as the magnon mode passes through the SAW frequency is clear. While we do observe a similar effect in Fig.3(d), the magnitude is much weaker which we believe is due to a combination of the mode drifting with time and the difficulty of positioning the YIG sphere at the focus, given the geometry shown in Fig.\ref{fig_setimg}(c).

 Our analysis here is complicated by the fact that there is significant electromagnetic crosstalk between the two ports and one needs to distinguish between the local magnon-SAW interaction occurring on the sample and possible interference effects occurring at the VNA. In particular, the electromagnetic radiation from the probes and the IDT, which act as inefficient antennas \cite{eggleston2015optical}, can excite the magnon mode and the scattered signal picked up by the receiving port will interfere with the acoustic transmission. We refer to this second pathway as a nonlocal interaction due to the phase sensitive detection employed by the VNA. In our experiments, the crosstalk signal is larger than the acoustic transmission because of the diffraction effects mentioned above. One can see this in action, by noting that in the case of $f_m>f_s$ and $f_m<f_s$, the magnon signal appears as a transmission peak rather than a dip, which is clear signature of a non-local interference pathway \cite{rieger2023fano}. One way to reduce this background is to exploit the significantly lower speed of sound compared to light and time-gate the VNA transmission \cite{wollensack2012vna}. The datasets shown in Fig.\ref{fig_data} have been time gated with a notch of 40 ns. While this helps to improve the signal to noise ratio (cf. the raw datasets in Appendix \ref{App: Raw_data_sets}), the high quality factor of the YIG sphere makes the cross talk persist for significantly longer than the electromagnetic transit time. This residual cross-talk makes it challenging for us to extract the SAW-magnon interaction strength from the experiments, although the excess local interaction with the SAW through the surface $B_1$ field is clear from Fig.\ref{fig_data}(a,b). As a control, we repeat the experiment with a straight IDT device and do not observe an excess attenuation at the magnon-SAW crossing (cf. Appendix \ref{App: Straight IDT}).

Given that the SAW-magnon mode interaction is observed with the YIG sphere not touching the sample, this experiment provides strong preliminary evidence that GHz frequency, localized $B_1$ fields can be generated on the surface of piezoelectric devices and can be used to interface with spin systems. While this is encouraging, we would like to emphasize that the experimental modality has a few limitations obvious with hindsight and the results should be interpreted within these constraints. In particular, the size of commercially available YIG spheres limits the sensitivity of the experiment by physically requiring the IDTs be separated by $>100\lambda_a$, and making it challenging to determine the focus in real-time. The size of the YIG sphere is also responsible for the magnitude of the crosstalk, which makes it challenging to infer the coupling strength and the coupling dynamics from our experiments. In particular, one of the key effects we were hoping to confirm was the frequency dependence of the interaction, which should scale $\propto f^2$. Although we made devices with varying SAW frequency, we were unable to quantify this effect. Moving forward, by impedance matching the transducers (to reduce electromagnetic radiation and reflections) and working with high Q YIG samples with dimensions $<$ \qty{5}{\um}, one can potentially scan the surface and verify the spatial extent of the $B_1$ field both in-plane ($x,y$) and in $z$. Mapping the spatial confinement of the $B_1$ field is critical to the spin sensitivity enhancement experiments and this is something we are unable to do with our current setup. Finally, moving to higher acoustic frequencies (and higher $B_z$) would enable us to saturate the YIG sphere and avoid the relaxation drift with time, which is another source of error in our experiments. In passing, we would like to note that the $B_1$ field description provides a complimentary route towards understanding the spin rotation effects in nanoparticles interacting with SAWs \cite{chudnovsky2016manipulating}, and interpret the switching of magnetization observed in previous experiments \cite{tejada2017switching}.

\section{Prospects for single spin electrical readout}

As noted in a recent review \cite{blank2017recent}, the problem of improving spin detection sensitivity in electron spin resonance experiments boils down to focusing the magnetic field to deeply sub-wavelength geometries while maintaining high-Q. In effect, piezoelectric microresonators are ideal in that they naturally provide both strong confinement and high Q, and therefore provide a natural complement to traditional electromagnetic approaches \cite{abhyankar2022recent}. The key issue is whether the magnetic field strengths can be substantial in these devices. As we have shown above using both scaling arguments and proof-of-principle experiments, the surface current density has a ${\propto}f^2$ and can be further enhanced by working with stronger piezoelectric materials / orientations ($K_{eff}^2>$  0.2) and designing small mode volume acoustic cavities \cite{msall2020focusing} to exploit the power scaling. With advances in materials and device geometries pushing acoustic device operation to ever-higher frequencies ($>$ \qty{50}{GHz}) \cite{xie2023sub}, the prospect of acoustics enabled X-Band ESR is within reach. There is still the open question of how one can efficiently load the near field of these devices efficiently to separate the pure field induced effects from strain effects. One possible route could be to employ suspended membranes (similar to \cite{teufel2011circuit}, but made with an insulator like alumina) in close proximity ($<$ \qty{50}{nm}) to the piezoelectric substrate.

We can estimate the minimum spin detection sensitivity, following \cite{bienfait2016reaching}:

\begin{equation}
N_{min} = \frac{\kappa}{2gp}\sqrt{\frac{nw}{\kappa_2}}
\end{equation}

where $N_{min}$ is the single-shot spin detection sensitivity (per echo), $\kappa$ is the total cavity decay rate given by $\kappa=\omega_0/Q_L$ with $\omega_0=2{\pi}f$ being the operating frequency and $Q_L$ the loaded Q factor of the cavity. We assume the cavity is operated at critical coupling with external coupling rate $\kappa_2\approx\kappa/2$, $n$ is the average number of noise photons, given by $n=k_BT/\hbar\omega_0$, $p$ is the spin polarization $p = \frac{1-e^{-\frac{\hbar\omega_0}{k_BT}}}{1+e^{-\frac{\hbar\omega_0}{k_BT}}}$, and the spin resonance width $w=2/T_2$ with $T_2$ the average spin dephasing time. $g$ is the spin cavity coupling rate defined above in eqn.\ref{eqn:g}. For operating frequencies of \qty{10}{GHz}, temperature \qty{4}{\kelvin}, mode volume \qty{5}{\um^3} and $T_2\approx$ \qty{50}{ms}, achieving $N_{min}\approx$ 1 requires $Q_L\approx10^5$, which is challenging for mechanical systems, but feasible given recent results \cite{bicer2023low} and the favourable scaling of acoustic dissipation with temperature. In any case, this shows that provided the $B_1$ field in piezoelectric microresonators has a spatial extent comparable to that of the acoustic field which requires experimental confirmation, then detecting individual magnetic nanoparticles at room temperature ($N_{spins}{\ge}10^5$) is well-within reach using current devices.

The exquisite spin detection potential of these resonators can be understood from a different perspective by treating them as the high frequency analogs of mechanical cantilever based spin sensors \cite{rugar2004single}, where the piezoelectric effect enables inductive detection. It was noted \cite{sidles1993signal} that the signal to noise ratio for both inductive and mechanical detection of spin resonance scales $\propto\sqrt{\omega_0Q/k_m}$, with $\omega_0$ is the operating frequency, $Q$ the cavity quality factor and $k_m$ an effective magnetic spring constant which scales with the cavity mode volume $V_m$ for inductive detection. The sensitivity enhancement therefore derives from achieving high quality factors in deeply sub-wavelength mode volumes, which gives an effective Purcell enhancement $\propto\sqrt{Q/V_m}$ to the sensitivity \cite{haroche2006exploring}.

We would like to conclude by noting that while in this work, we have primarily focused on using piezoelectric devices for improving spin detection sensitivity in ESR, the strong field confinement is also of interest in scenarios involving large-scale closely packed efficient spin addressing \cite{weng2024crosstalk} without deleterious crosstalk effects, as would be necessary in future spin-based quantum computing.

\section{Acknowledgements}

We would like to thank John Rarity, Joe Smith, Vivek Tiwari, Hao-Cheng Weng, Alex Clark, Rowan Hoggarth and Edmund Harbord for valuable discussions and suggestions. We acknowledge funding support from the UK's Engineering and Physical Sciences Research Council (EP/N015126/1, EP/V048856/1) and the European Research Council (ERC-StG SBS 3-5, 758843).

\begin{appendix}

\section{Spin Momentum Locking of Evanescent fields in surface acoustic waves and surface plasmons}
    \label{App:SAW_SPP}

\begin{figure} [!htbp]
    \centering
    \includegraphics[width = 1.0 \columnwidth]{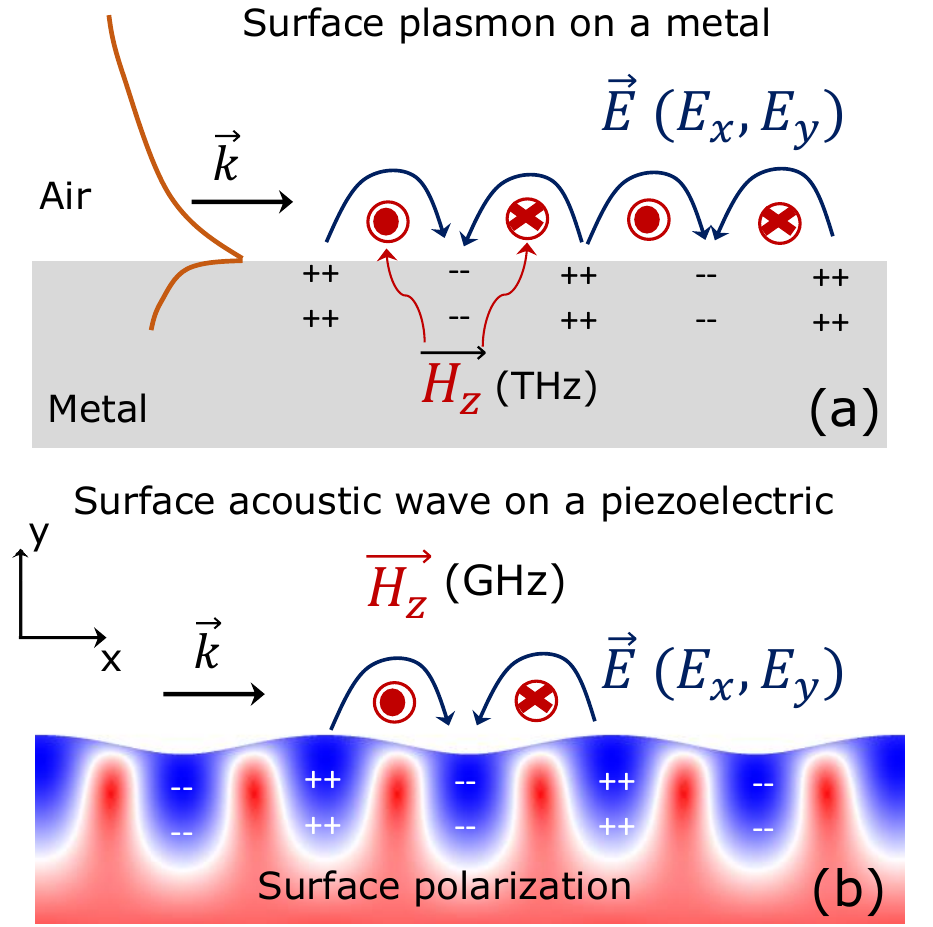}
    \caption{(a) Schematic illustration of surface plasmon propagating on a metal-air interface. The evanescent electric field and the magnetic field orientation are shown, alongwith the surface charges that terminat the electric field on the metal surface. (b) Illustration of a surface acoustic wave propagating on a piezoelectric material. The evanescent electric field orientation is similar to that in (a), but terminated by bound polarization charges induced in the piezoelectric. By analogy, the orientation of the $B_1$ field can be determined, as shown.}
    \label{fig_SAW_SPP}
\end{figure}

One can see the universality of spin-momentum locking \cite{van2016universal} as applied to all evanescent fields by looking at two very different surface waves: a surface plasmon propagating at a metal air interface (at $>$\qty{100}{THz}) and a surface acoustic wave ($<$ \qty{50}{GHz}) propagating at a piezoelectric-air interface. The respective cases are shown in Fig.\ref{fig_SAW_SPP}(a,b). As can be seen in both cases, the evanescent electric fields curl with an orientation determined by spin-momentum locking. The main difference between the two scenarios is the termination of the fields on free charges in the metal and on bound polarization charges in the piezoelectric case. Given that the surface plasmon dispersion relation is traditionally derived by solving for the magnetic field, the $B_1$ orientation can be derived by analogy as shown in Fig.\ref{fig_SAW_SPP}(b).   

\section{Uncorrected datasets without time-gating}
    \label{App: Raw_data_sets}
\begin{figure}[h!]
    \centering
    \includegraphics[width = 1.0 \columnwidth]{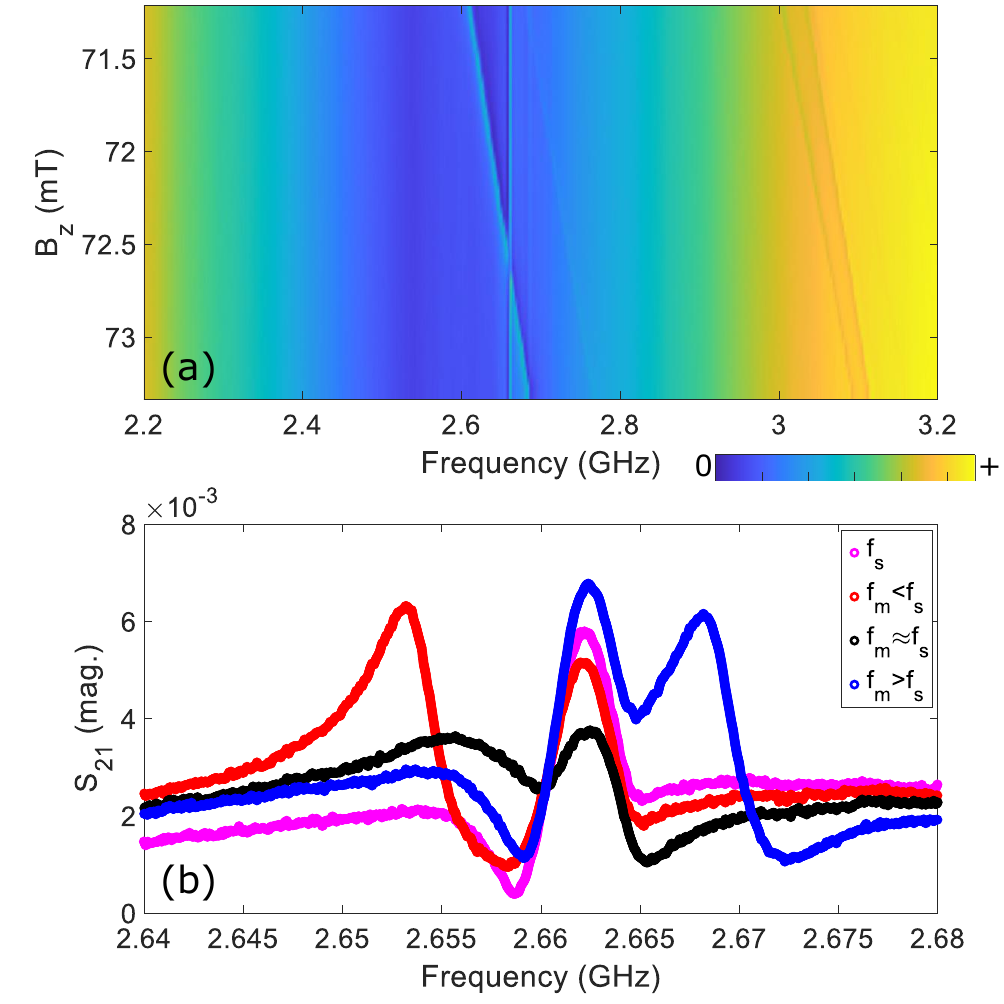}
    \caption{(a) Uncorrected data sets for the data corresponding to Fig.\ref{fig_data}(a,b). Without the time-gating, the background electromagnetic crosstalk makes it impossible to observe the excess attenuation during the SAW-magnon mode crossing, although even with the raw dataset, a net attenuation (cf. black curve) is clearly visible. }
    \label{fig_rawdata}
\end{figure} 

Figure \ref{fig_rawdata} plots the raw data sets corresponding to the time-gated datasets plotted in Fig.\ref{fig_data}(a,b). As discussed in the main text, the presence of the background electromagnetic crosstalk makes it challenging to infer the SAW-magnon interaction, which can be best seen by comparing the black curves ($f_m{\approx}f_s$) in Fig.\ref{fig_data}(b) and Fig.\ref{fig_rawdata}(b). On the other hand, even with the raw datasets, we can clearly observe an overall attenuation as the magnon mode frequency comes close to the SAW frequency. This is in contrast to what we observe with the straight IDT devices, as discussed in the next section.

We would like to note that the choice of the notch gate time (\qty{40}{ns}) was not optimized for this analysis and neither was the gate shape. Given the separation between IDTs was \qty{200}{\um}, and a speed of sound of $\approx$ \qty{3750}{\meter\per\sec}, the acoustic wave takes $\approx$ \qty{53}{ns} to reach the receiving IDT. The choice of gate time was made as a rough trade-off between minimizing the background crosstalk signal and ensuring minimal attenuation of the acoustic signal in the transmitted spectrum. We would like to note again that time gating does not fully eliminate the background crosstalk because of the Q factor of the YIG sphere.

\section{Straight IDT control results}
    \label{App: Straight IDT}
\begin{figure}[!h]
    \centering
    \includegraphics[width = 1.0 \columnwidth]{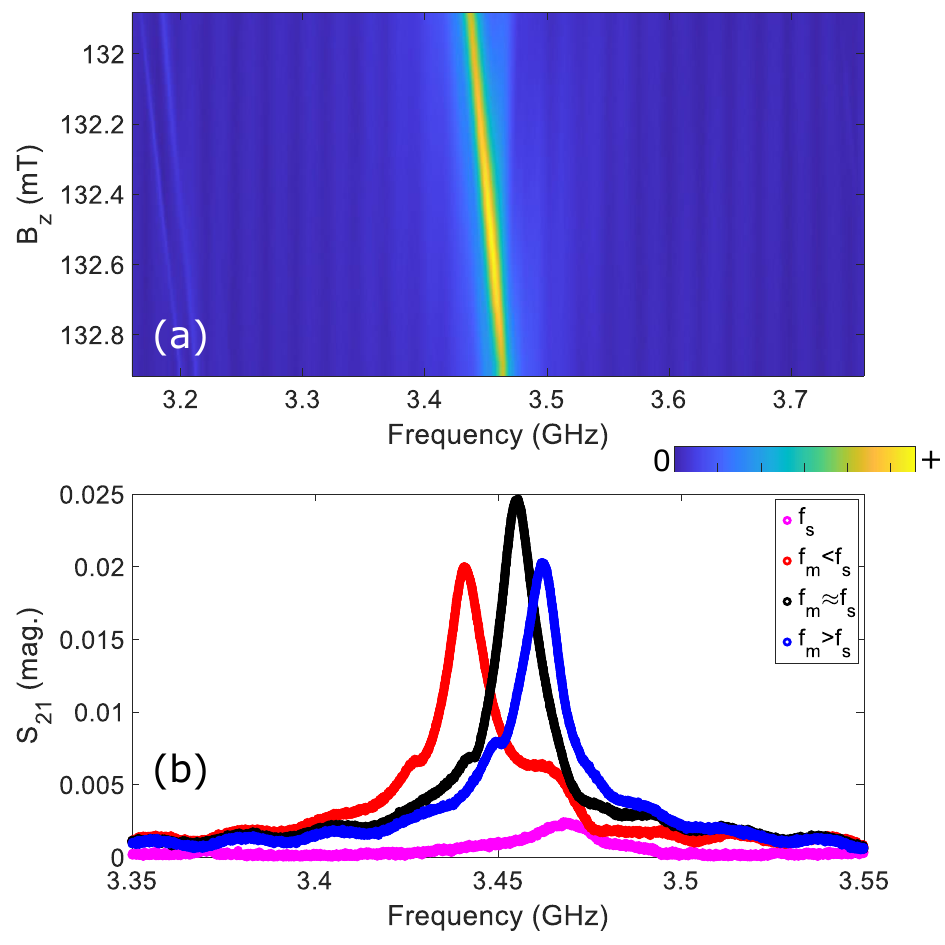}
    \caption{(a) 2D colorplot of the time-gated $\lvert{S_{21}}\rvert$ for a straight IDT transmit-receive device with the YIG sphere positioned in between. The YIG was mounted vertically in this experiment to align the 110 axis with $z$, which shifts the modes to higher frequencies in the \qty{3.4}{GHz} range (b) 1D datasets from (a) corresponding to the three different cases ($f_m<f_s$, $f_m{\approx}f_s$, and $f_m>f_s$), along with the background SAW transmission (magenta). The excess attenuation at the magnon-SAW crossing is not observable here.}
    \label{fig_stidt}
\end{figure} 

The power dependence of $J^2$ in equation \ref{eq:J_f} can be tested by measuring the relative performance of IDT with curved and straight fingers in inducing magnon absorption. Given the local field intensity in curved devices at the focus is increased by the focusing ratio, the straight IDT devices here serve as a control experiment to separate the local SAW-magnon interaction from the nonlocal EM crosstalk-SAW interaction occuring in the VNA, discussed in the main text. 

Experiments identical to that reported in Fig.\ref{fig_data} were done with a straight IDT device. The YIG sphere in these experiments was mounted vertically with a view towards saturating the sphere and avoiding the time drift in the magnon modes. This moves the magnon modes to higher frequencies in the 3.4 GHz range, and the IDT period was reduced correspondingly to shift the SAW response to higher frequencies. Fig.\ref{fig_stidt}(a) plots the 2D colorplot of the transmitted ${\lvert}S_{21}{\rvert}$ as a function of frequency and $B_z$. To achieve the higher $B_z$, we use a permanent magnet in combination with our electromagnet. Fig.\ref{fig_stidt}(b) shows linecuts from \ref{fig_stidt}(a) that correspond to the three different cases $f_m<f_s$, $f_m{\approx}f_s$ and $f_m>f_s$. Here, we don't observe an excess attenuation as the magnon mode crosses through the SAW resonance, which can be interpreted as a signature of the dependence of the local current density on the local power density. 

While mounting the YIG sphere vertically makes the magnon mode frequency stable due to field saturation, it makes it very challenging to determine the positioning of the sphere with respect to the beam focus using imaging cameras. In particular, the top view camera, shown in Fig.\ref{fig_setimg}(a) can not be used anymore and one has to rely more on the side view camera with associated parallax errors. We tried to repeat the experiments in Fig.\ref{fig_data} with focusing IDTs and the sphere mounted vertically, but we were not able to observe a clear signature of excess attenuation as in Fig.\ref{fig_data}(b,d) which we currently attribute to the difficulty of positioning the sphere at the beam focus in this configuration. As noted in the main text, many of the issues detailed here can be addressed by moving to high Q YIG samples with dimensions $<$\qty{5}{\um}.

\section{Reference IDT spectra}
    \label{App: IDT_spectra}

\begin{figure}[!h]
    \centering
    \includegraphics[width = 1.0 \columnwidth]{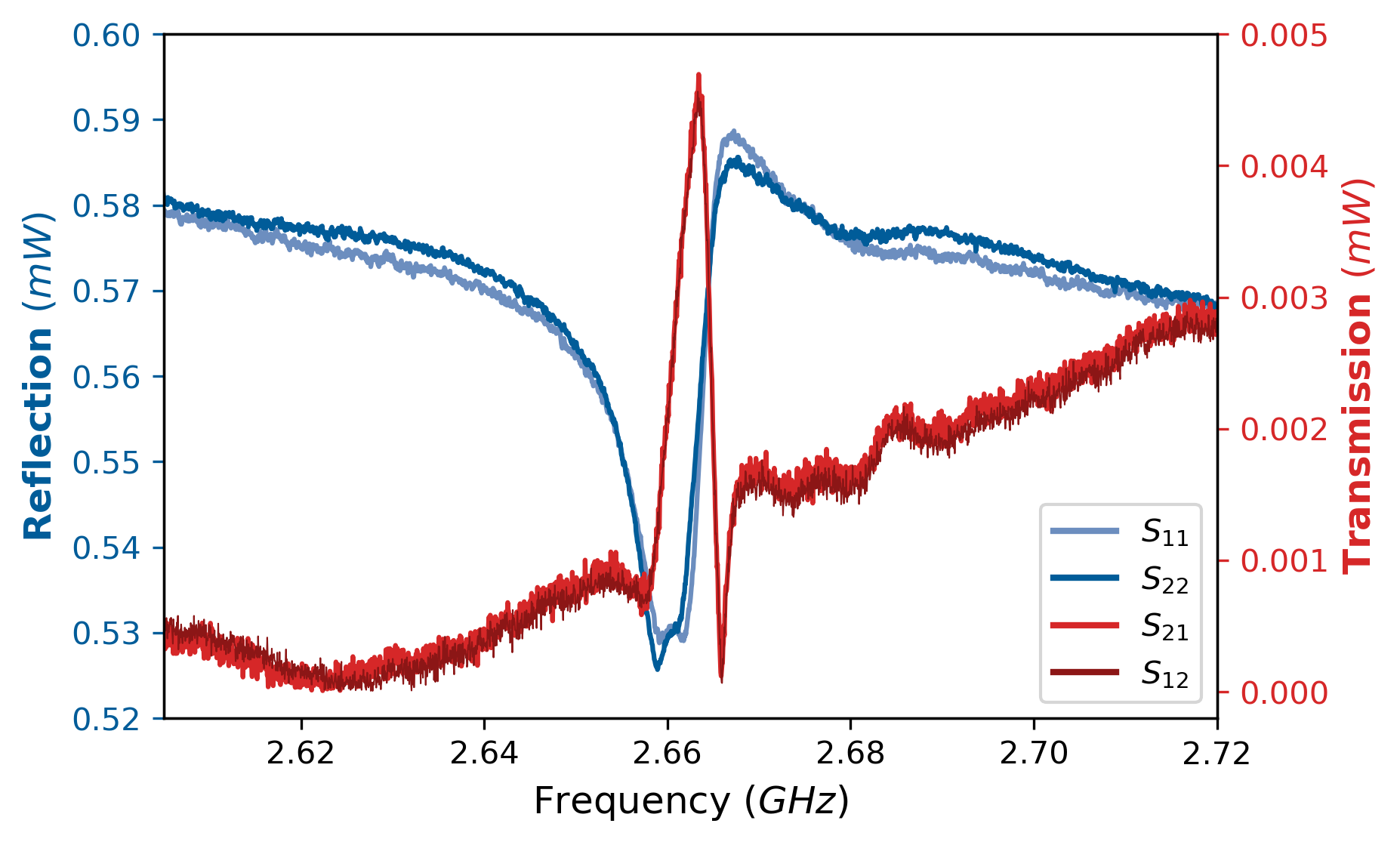}
    \caption{Bare IDT (without the YIG sphere) reflection ($S_{11}$, $S_{21}$) and transmission ($S_{12}$, $S_{21}$) spectra for the device in Fig.\ref{fig_data}(a,b) with VNA power \qty{1}{mW}). The spectra correspond to the raw VNA measurements without time-gating.}
    \label{fig_idtref}.  
\end{figure} 

\end{appendix}

\clearpage
\bibliography{References}

\end{document}